# Microcontroller Based Load Monitoring System

Madhavanunni A. N.[1], Arya S. S.[2], Renjith Kumar D.[3]

*Department of Electrical & Electronics Engineering*
*College of Engineering Adoor, India*
[1]madhavanunni77@gmail.com, [2]ssaru.sura@gmail.com, [3]renjithkumarcea@gmail.com

*Abstract* – **The demand for power has increased exponentially over the last century. One avenue through which today's energy problems can be addressed is through the reduction of energy usage in households. This has increased the emphasis on the need for accurate and economic methods of power measurement. The goal of providing such data is to optimize and reduce their power consumption. In view of this, the present manuscript focuses on the design and implementation of precise and reliable load monitoring system using PIC microcontroller chip (PIC16F877A). This involves an accurate sensing of voltage, current and power factor of the load. A clever utilization of in-built ADC and timers of the microcontroller reduces the design complexity of the system. The proposed system monitors the load continuously on a real time basis and displays the parameters such as voltage, current, power factor, active, reactive and apparent powers in an LCD module. The use of microcontroller reduces the cost and makes the device compact. The proposed system has been implemented and tested in the laboratory for single phase loads.**

## I. INTRODUCTION

Measurement plays a significant role in achieving goals and objectives of engineering because of the feedback information supplied by them [1]. The advancement of science and technology is dependent upon a parallel progress in measurement techniques. As per the modern measuring instrument requirements aiming high sensitivity, faster response, greater flexibility, low cost and less time consumption there has been a tremendous intrusion of electronic components in this area. Among them the microcontrollers play a vital role.

In the present day scenario, load monitoring is much essential and plays a vital role in the field of power systems because it provides the optimized information about the performance of the load. An efficient monitoring system that provides much accurate measurement values has a great impact on the realization of automation of power network management and the safety and economic operation of the power network. Several measurement techniques have been developed so far. The digital measurement systems have gained upper hand over the conventional analog meters because of its accuracy and high speed measurement characteristics. Microcontrollers, DSP (Digital Signal Processor) and ARM (Advanced RISC Machines) are extensively utilized for the accurate and high speed measurement of power parameters [2]-[6]. DSP and ARM based digital measurement systems make use of the sampling technique to derive an accurate measurement system [3], [4]. Microcontrollers like Peripheral Interface Controller (PIC 16F877A) and single chip microcontroller (C8051F020) are widely used for measurement purposes as it reduces the circuit complexity and provides an appreciable accuracy with a minimum possible error [7]-[8]. Modern digital measurement techniques include the generation of Pulse Width Modulated (PWM) signals and various microcontroller based circuits to measure the power factor and other power parameters [9]-[11].

This paper focuses on the design and implementation of a precise and reliable load monitoring system using PIC microcontroller chip in which voltage, current, power factor, active, reactive and apparent powers are displayed in a single LCD screen. The effective power measurement facilitates obtaining the most precise feedback regarding the state of an electrical load. As far as modern methods are concerned, they have been successfully implemented for a wide range of applications yet they suffer from a little bit of complex design circuitry. As the proposed circuit is based on programming, component and device losses inside the circuit are much less as compared to the other designs thus increasing the efficiency and reliability.

## II. PROPOSED SYSTEM

Microcontroller based load monitoring system is shown in Fig. 1.

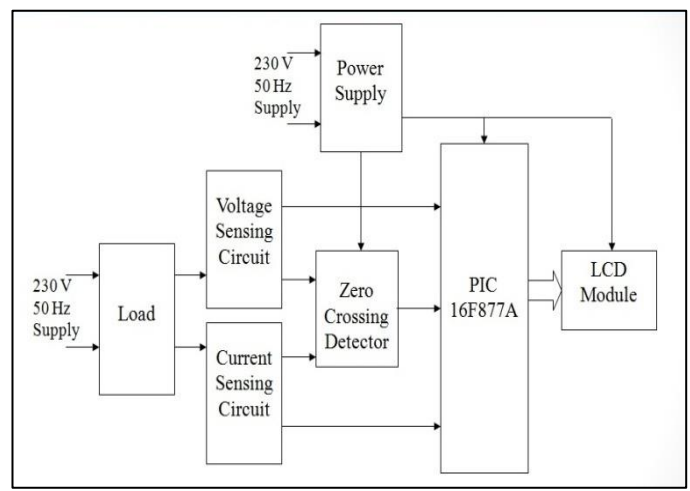

Fig. 1 Block diagram of the proposed system

The principal element in the circuit is PIC microcontroller (16F877A) that works with 10 MHz crystal in this scheme. The current and voltage signals are acquired from the load by using current and voltage transducers. These acquired signals are then passed on to the zero crossing detector IC individually that transpose both the current and voltage waveforms to square-waves to make perceivable for the microcontroller to observe the zero crossing of current and voltage signals. Microcontroller reads the RMS values of voltage, current, power factor, active, reactive and apparent powers using the algorithm written in it. The load is being monitored continuously and the measured parameters are displayed on the LCD screen. The proposed system is designed for monitoring the loads with the ratings up to 250V, 20A.

*A. Current and Voltage Sensing*

Current sensing and voltage sensing circuits are used to acquire the load current and voltage signals for the calculation of current, voltage and power factor. Current and voltage signals are acquired from the load and stepped down to safer values using transducers and voltage divider networks. These signals are then applied to the zero crossing detector and to the microcontroller for further calculations.

*B. Zero Crossing Detection*

Zero crossing detection is the most common method for measuring the phase and frequency of a periodic signal. Zero crossing is the point of choice for measurement of phase [12]. The zero crossing detector converts both current and voltage waveforms to square-waves to make perceivable for the microcontroller to detect the delay between both the signals at the same time instant. It helps in calculating the load power factor.

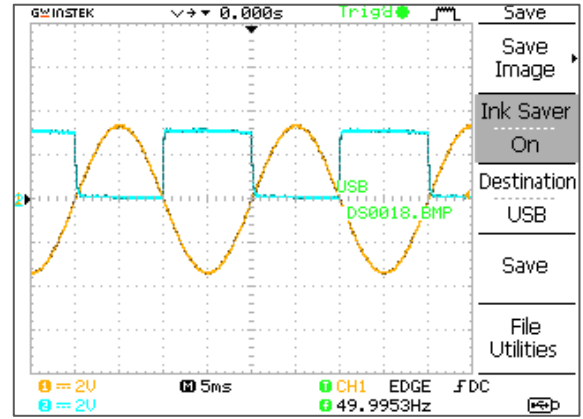

Fig. 3 Input and output waveforms of ZCD

III. CALCULATION OF POWER PARAMETERS

The calculation of power parameters is carried out by the microcontroller using the following equations [13].

$$\text{Active power, } P = VI\cos\theta \quad (1)$$

$$\text{Reactive power, } Q = VI\sin\theta \quad (2)$$

$$\text{Apparent power, } S = \sqrt{(P^2 + Q^2)} = VI \quad (3)$$

where, $V$ is the load voltage, $I$ is the load current and $\theta$ is the phase angle between the load voltage and load current.

IV. MICROCONTROLLER ALGORITHM SCHEME

Fig. 4 shows the flow chart of the proposed load monitoring system.

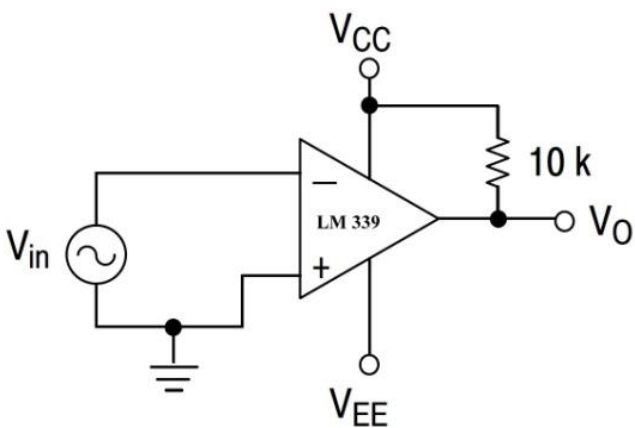

Fig. 2 Zero crossing detector

Fig. 2 shows the basic zero crossing detector and Fig. 3 shows its input and output waveforms.

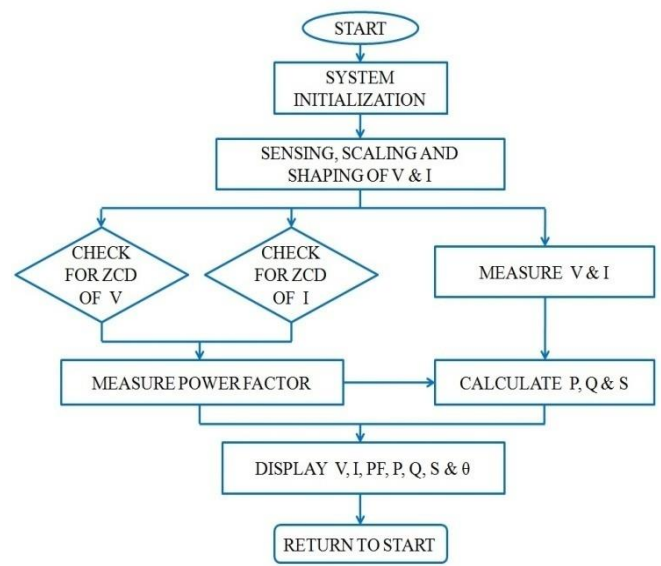

Fig. 4 Flowchart of the proposed system

In the flow chart, the first step is about initializing the system. Microcontroller measures the load voltage and load current on a real time basis using the inbuilt ADC module. The voltage and current signals which have been converted into pulses after zero crossing are provided to microcontroller input pins (CCP1 and CCP2) that are fundamentally the input of capture modules of the microcontroller.

### V. SIMULATION RESULTS AND DISCUSSIONS

The proposed system is completely tested on Proteus Professional in which simulation results are based on the lagging power factor of the load. Following are the simulation results which include different types of loads.

*a. Case 1: For resistive load*

For a resistive load, the current and voltage signals are in phase and the corresponding measurements are as shown in Fig. 5.

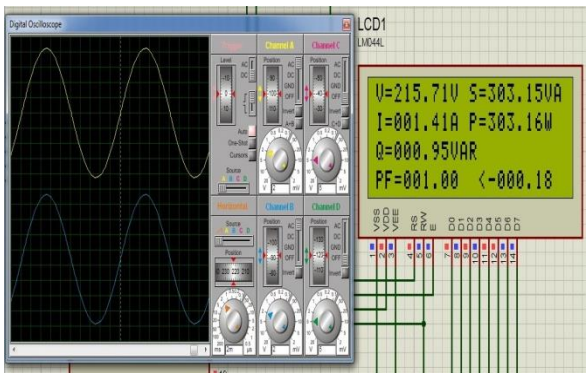

Fig. 5 Simulation results for resistive load

*b. Case 2: For inductive load*

For an inductive load, the current lags behind the voltage as shown in Fig. 6 and so the load consumes an appreciable reactive power depending upon its power factor.

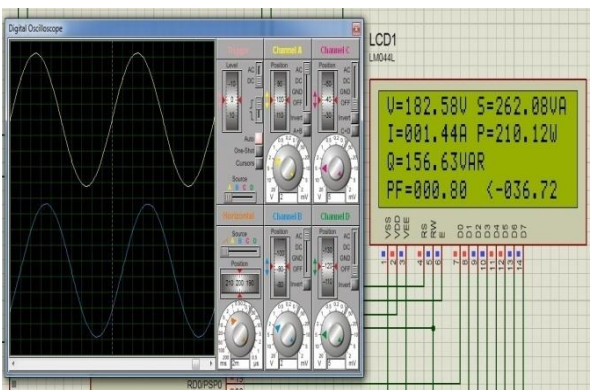

Fig. 6 Simulation results for inductive load

### VI. HARDWARE RESULTS AND DISCUSSIONS

The proposed system is implemented and tested in the laboratory for resistive and inductive loads. The following are the hardware results obtained.

*a. Case 1: For resistive load*

In this case, the current and voltage signals are in phase as shown in Fig. 7. The zero crossing detection of current and voltage signals for a resistive load is shown in Fig. 8 and corresponding load monitoring is shown in Fig. 9.

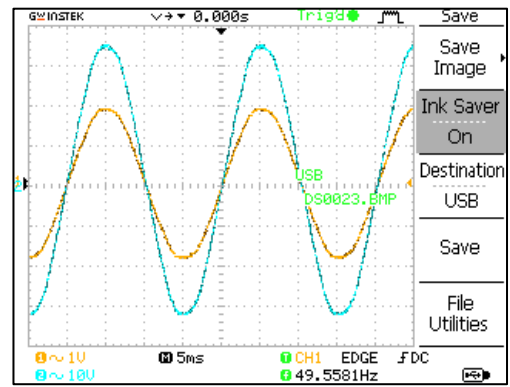

Fig. 7 Current and voltage signals for resistive load

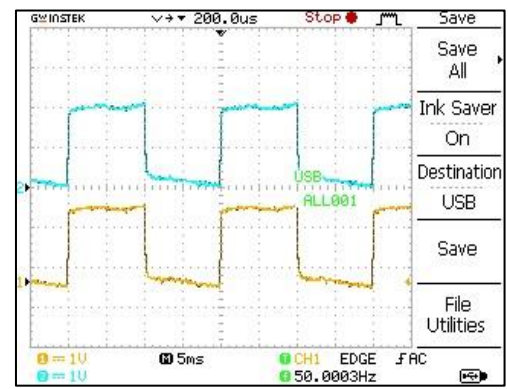

Fig. 8 Zero crossing detection for resistive load

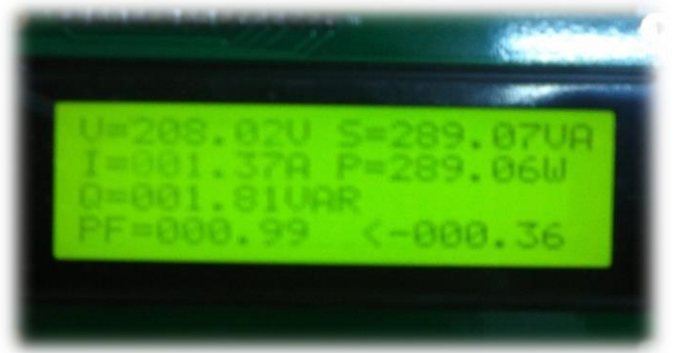

Fig. 9 Load monitoring for resistive load

*b. Case 2: For inductive load (0.8 pf)*

For an inductive load, the current lags behind the voltage and results in low power factor. The load current and voltage signals are depicted in Fig. 10. The corresponding zero crossing detection of those signals is depicted in Fig. 11 and Fig. 12 shows the load monitoring for the inductive load.

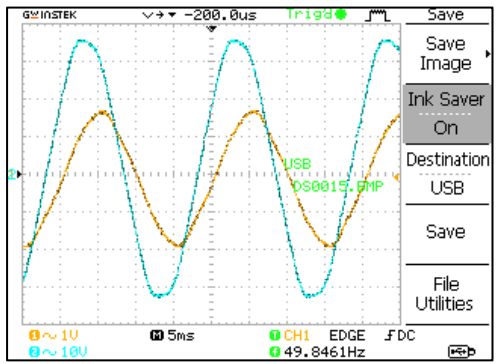

Fig. 10 Current and voltage signals for 0.8 pf inductive load

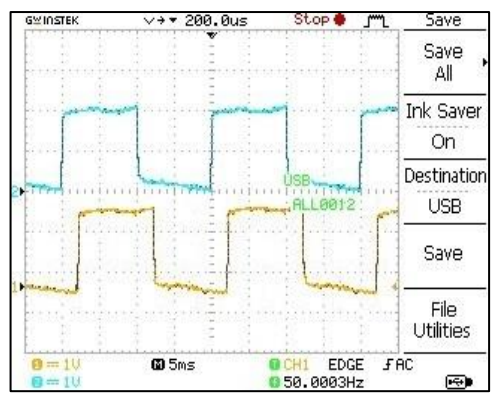

Fig. 11 Zero crossing detection for 0.8 pf inductive load

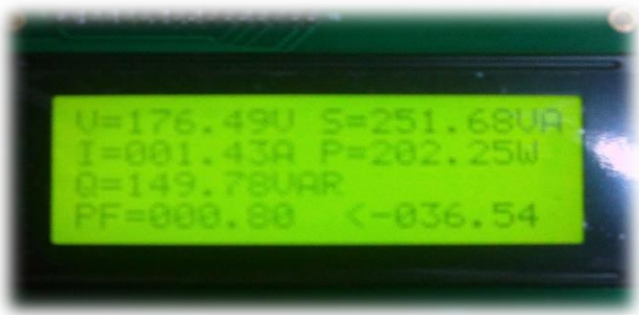

Fig. 12 Load monitoring for 0.8 pf inductive load

The system is more accurate than the analog meters which uses conventional methods for the measurement of these parameters. Moreover the proposed system is simple and economic.

## VII. CONCLUSIONS

Harmonic loss reduction and improvement in transmission efficiency are the prominent challenges of power systems. This has increased the emphasis on the need for accurate and economic methods of power measurement in recent years. The above presented manuscript focuses on the design and implementation of a system which monitors the load continuously on the real time basis using PIC 16F877A. The designed system is simple and less expensive with an accuracy of two decimal points. The proposed system can be enhanced to operate in other frequencies too by introducing a Phase Locked Loop (PLL) mechanism. The extended form can be obtained for 3-phase loads also. The system may be further enhanced to incorporate design implementation and testing on nonlinear loads. The future developments include the designing of reactive power compensation system that works on the basis of the feedback provided by this load monitoring system.

### ACKNOWLEDGEMENT

The authors are really thankful to the Department of Electrical & Electronics Engineering, College of Engineering Adoor for providing the technical assistance and support without which it was difficult to implement the system.